\documentclass[aps,pra,onecolumn,amsmath,amssymb]{revtex4}

\usepackage{graphicx}
\usepackage{dcolumn}
\usepackage{bm}
\usepackage{epsfig}
\usepackage{epsfig}
\usepackage{amsmath}
\usepackage{amsfonts}

\newcommand{\beq}{\begin{equation}}
\newcommand{\eeq}{\end{equation}}
\newcommand{\bea}{\begin{eqnarray}}
\newcommand{\eea}{\end{eqnarray}}

\usepackage{theorem}

\theorembodyfont{\rmfamily}

\theoremstyle{break}

\def\QED{~\rule[-1pt]{5pt}{5pt}\par\medskip}

\begin{document}


\title{Reduced coupling with global pulses in quantum registers }

\author{Haidong Yuan$^1$,Van D. M. Koroleva$^2$,Navin Khaneja$^2$}
\address{$^1$Department of Applied Mathematics, The Hong Kong Polytechnic University, Hong Kong}
\address{$^2$School of Engineering and Applied Sciences, Harvard University,
33 Oxford Street, Cambridge MA 02138}
\email{haidong.yuan@gmail.com,do.maivan@gmail.com}

\date{\today}

\begin{abstract}
Decoupling is an important tool to prolong the coherence time of quantum systems. Most decoupling schemes have been assuming selective controls on the system and it is believed that with global pulses one can only decouple systems with certain coupling terms like secular dipole-dipole coupling. In this article we show that with global pulses it is possible to reduce the coupling strength of other types of coupling, which we demonstrate with Ising coupling. The complexity of such pulses is independent of the size of system.
\end{abstract}

\maketitle

\section{\label{sec:introduction}Introduction}
Quantum systems suffer from decoherence due to interactions with environments. The task of decoupling is to remove unwanted couplings between systems and environments~\cite{Viola1998, Viola1999, Viola19992}. Many decoupling schemes have been proposed and demonstrated in experiments~\cite{Uhrig,Biercuk2009,Biercuk20092,Khodja2005,Khodja2007,QDD,Ryan2010,Viola2005,Wocjan,Viola2006,Yang2008,Uhrig2010,Jenista2009,Szwer2011,Ajoy2011,Wang2011}, for example, randomized dynamical decoupling~\cite{Viola2005} uses randomly selected pulses at regular intervals, UDD (Uhrig dynamical decoupling)~\cite{Uhrig} can cancel dephasing of a single qubit up to order $n$ by using a minimal number of $n$ pulses, CDD (concatenated dynamical decoupling) constructs decoupling pulse sequences recursively\cite{Khodja2007}. There are also studies on using pulses to remove internal couplings of quantum systems~\cite{Ernst,Solid,Wocjan} or engineer Hamiltonians~\cite{Borneman2012,Bookatz2013}.

A common feature of these decoupling schemes is that they all assume selective controls on the system. For many quantum systems, selective addressing of each qubit could be very high demanding, especially of those systems whose environment consists of the same physical objects as the system,for example, in some solid state devices, the system and the environment can be the spin of same nuclear species. For such systems, selective controls on the system is very hard, as the pulses usually affect all the spins, i.e., the pulses will be global. The known examples of decoupling with global pulses are WAHUHA~\cite{WAHUHA}, MREV-8 and MREV-16~\cite{MREV1,MREV2} in Nuclear Magnetic Resonance, which exploit the symmetry of homonuclear secular dipole-dipole coupling to decouple the system. Such decoupling schemes rely on the symmetry of secular dipole-dipole coupling and therefore do not apply to other types of coupling. Recently applications of global pulses in Hamiltonian engineering were also studied\cite{Ajoy2013}. In this paper, we examine the use of global pulses to decouple the system with Ising coupling between the qubits, and show that, to the contrary of previous belief it is possible to decouple system with couplings different from secular dipole-dipole coupling. The advantage of global pulses is that the number of pulses needed for decoupling will be independent of the number of qubits, i.e. the complexity of global pulses is $O(1)$.

\section{Average Hamiltonian}\label{sec:basic}
The principle of decoupling can be illustrated by the average Hamiltonian theory\cite{Magnus1954,AHT}, i.e., the propagator can be written as a single exponential relying on some Average Hamiltonian $\overline{H}$ which has the same effect as a time varying Hamiltonian. The full advantage of this theory is often realized in an interaction frame of a \textit{period} and \textit{cyclic} Hamiltonian. Assume that in an appropriate interaction frame, the Hamiltonian is piecewise constant $H_1, H_2, \ldots H_m$ in corresponding time intervals $t_1, t_2, \ldots, t_m$, then $$e^{-i\overline{H}t}=e^{-iH_mt_m}\cdots e^{-iH_1t_1}.$$

Here $H_1\cdots H_m$ are transformed Hamiltonians from the physical Hamiltonian by applying pulses on the system, i.e., $H_i=U_i^\dagger H U_i$ where $U_i$ represents the propagator generated by the pulses. The first few orders of the average Hamiltonian are
$$\overline{H}=\overline{H}^{(0)}+\overline{H}^{(1)}+\overline{H}^{(2)}\cdots,$$
\bea
\aligned
\overline{H}^{(0)}&=\frac{1}{t}(H_1t_1+H_2t_2+\cdots+H_m t_m),\\
\overline{H}^{(1)}&=-\frac{i}{2t}\{[H_2t_2,H_1t_1]+[H_3t_3,H_1t_1]\\
&+[H_3t_3,H_2t_2]+\cdots\},\\
\overline{H}^{(2)}&=\frac{1}{12t}\{[H_2t_2,[H_2t_2,H_1t_1]]-[H_1t_1,[H_2t_2,H_1t_1]]+\cdots\}
\endaligned
\eea
where $t=\sum_{i=1}^mt_i.$

A global pulse transforms an initial Hamiltonian $H$ to $$H_i=(U\otimes U\otimes\cdots \otimes U)^\dagger H U\otimes U\otimes\cdots \otimes U,$$ where $U\in SU(2)$ represents the propagator generated by the global pulse on each qubit. Global pulses have been used to decouple systems with secular dipole-dipole coupling. In this case, the coupling Hamiltonian takes the following form,
\beq\label{ddHami1}
H_1^{dd}=\sum_{jk}d^{jk}(2I_{jz}I_{kz}-I_{jx}I_{kx}-I_{jy}I_{ky}),
\eeq
here $I_{x}:= \frac{1}{2}\left(\begin{smallmatrix}
0 & 1 \\
1 & 0
\end{smallmatrix}
\right)$, $ I_{y}:= \frac{1}{2}\left(
\begin{smallmatrix}
0 & -i \\
i & 0
\end{smallmatrix}
\right) $, and $ I_{z}:= \frac{1}{2}\left(
\begin{smallmatrix}
1 & 0 \\
0 & -1
\end{smallmatrix}
\right) $, are the Pauli spin matrices and we denote $I_{\ell \nu}$ as the operator that
acts as $I_{\nu}$ on the $\ell$th spin(see~\cite{Ernst}).

Applying global $(x)_{\frac{\pi}{2}}$ pulse and $(y)_{\frac{\pi}{2}}$ pulse on the system, one gets
\bea\label{ddHamil2}
\aligned
H_2^{dd}=&\sum_{jk}d^{jk}(2I_{jy}I_{ky}-I_{jx}I_{kx}-I_{jz}I_{kz}),\\
H_3^{dd}=&\sum_{jk}d^{jk}(2I_{jx}I_{kx}-I_{jz}I_{kz}-I_{jy}I_{ky}).
\endaligned
\eea

It is easy to see that averaging these three Hamiltonians decouples the dipole-dipole coupling, i.e., $$e^{-iH_1^{dd}\delta t}e^{-iH_2^{dd}\delta t}e^{-iH_3^{dd}\delta t}=e^{-i3\delta t\overline{H}},$$ where $\overline{H}=0$. This is the basic building block of WAHUHA, MREV-8 and MREV-16~\cite{Solid} and the effectiveness of such decoupling scheme has been experimentally demonstrated~\cite{Suter2006}. However such decoupling scheme only works for the systems with coupling $\alpha I_{x}I_{x}+\beta I_{y}I_{y}+\gamma I_{z}I_{z}$ where $\alpha+\beta+\gamma=0$.

\section{Reduce Ising coupling with global pulses}\label{sec:Ising}
At a first look, it may seem impossible to decouple Ising coupling with global pulses, as global pulses can not change the signs of Ising coupling. We will show that it is indeed possible by extending our previous study on Homonuclear decoupling~\cite{Van2012}.

\begin{figure}[bt]
\begin{center}
\includegraphics[scale=.5]{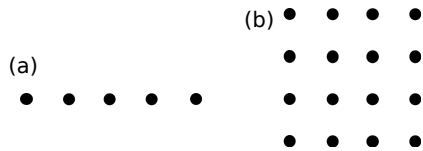}
\end{center}
\caption{Spin topology: (a)Linear Chain (b)Square Lattice }
\label{fig:topology}
\end{figure}

Consider a system consisting of $N$ qubits connected by Ising coupling, the coupling topology can take various shapes, for example, it can be a spin chain or spin lattice, as shown in Fig.~(\ref{fig:topology}).

A gradient magnetic field is added upon the system which induces Zeeman splitting on the qubits. The magnetic field and its gradient are large enough so that the differences of Zeeman splitting between coupled qubits are much larger than the coupling strength between the qubits. Under such gradient magnetic field, the Hamiltonian of the system takes the form
\beq
\label{eq:Hamiltonian}
H=\sum_{j=1}^{N}\omega_j I_{jz}+\sum_{(jk)\in G}J_{jk}I_{jz} I_{kz}
\eeq
where $\omega_j=-\vec{\mu}\cdot \vec{B_j}$, $\vec{\mu}$ is the magnetic moment and $B_j$ indicates the magnetic field at site $j$. $G$ is a graph indicating the coupling topology of the system, i.e., if the edge $(jk)\in G$, then the qubits at site $j$ and $k$ are coupled. We assume $\omega_i\gg J_{jk}$, $|\omega_j-\omega_k|\gg J_{jk},\forall i,j,k$.

We will first use a two-qubit system to illustrate the decoupling strategy, then generalize it to $N$ qubits with various coupling topologies.

For a two-qubit system, the Hamiltonian is
\beq
H_0=\omega_1 I_{1z}+\omega_2 I_{2z}+JI_{1z} I_{2z}.
\eeq

The basic pulse sequence consists of four periods, within each period it evolves according to the following Hamiltonians
\beq
\label{eq:basicstep}
\begin{split}
H_1&=\omega_1 I_{1z}+\omega_2 I_{2z}+JI_{1z} I_{2z}+A(I_{1x}+I_{2x}),\\
H_2&=-\omega_1 I_{1z}-\omega_2 I_{2z}+JI_{1z} I_{2z}+A(I_{1x}+I_{2x}),\\
H_3&=-\omega_1 I_{1z}-\omega_2 I_{2z}+JI_{1z} I_{2z}-A(I_{1x}+I_{2x}),\\
H_4&=\omega_1 I_{1z}+\omega_2 I_{2z}+JI_{1z} I_{2z}-A(I_{1x}+I_{2x}),
\end{split}
\eeq
where $H_1$ is obtained simply by applying a magnetic field in $x$ direction with effective amplitude $A$, $H_2$ is obtained from $H_1$ by conjugating a $\pi$-pulse along the $x$ direction, i.e., $H_2=e^{i\pi(I_{1x}+I_{2x})}H_1 e^{-i\pi(I_{1x}+I_{2x})}$, $H_4$ and $H_3$ are obtained with a control field along the $-x$ direction and conjugation with $\pi$-pulses along $x$ direction.
The $\pi$-pulses here are assumed to be infinitely narrow pulses.

Each of the four Hamiltonian is maintained for a period of $\Delta t$, keep the terms up to the second order of the average Hamiltonian theory, we obtain the following average Hamiltonian over an interval of $4\Delta t$,
\beq
\label{eq:1eff}
\aligned
H_{1eff}&=J I_{1z}I_{2z}+\frac{A\Delta t}{2}(\omega_1 I_{1y}+\omega_2 I_{2y})+ A\Delta t J(I_{1y} I_{2z}+I_{1z} I_{2y})\\&+\frac{(A\Delta t)^2}{2}(\omega_1 I_{1z}+\omega_2 I_{2z}) +\frac{4}{3}(A\Delta t)^2 J(I_{1y} I_{2y}-I_{1z} I_{2z})+O((A\Delta t)^3)
\endaligned
\eeq

Denote $\theta=\Delta t A$ and choose $\Delta t$ such that $\theta \ll 1$, but $\theta|\omega_1-\omega_2|\gg J$.
Next we apply $\pi$ pulses along $y$ direction to flip the signs of the third and fourth terms in $H_{1eff}$. As a consequence, we create $H_{2eff}$ such that

\beq
e^{-iH_{2eff}8\Delta t}=
e^{-iH_{1eff}4\Delta t}e^{i\pi(I_{1y}+I_{2y})}e^{-iH_{1eff}4\Delta t} e^{-i\pi(I_{1y}+I_{2y})}.
\eeq

It is straightforward to see that
\beq
H_{2eff}=\frac{\theta}{2}(\omega_1 I_{1y}+\omega_2 I_{2y})+J I_z S_z+\frac{4}{3}\theta^2J(I_{1y} I_{2y}-I_{1z} I_{2z})+O(\theta^3).
\eeq

\begin{figure}[bt]
\begin{center}
\includegraphics[width=\columnwidth]{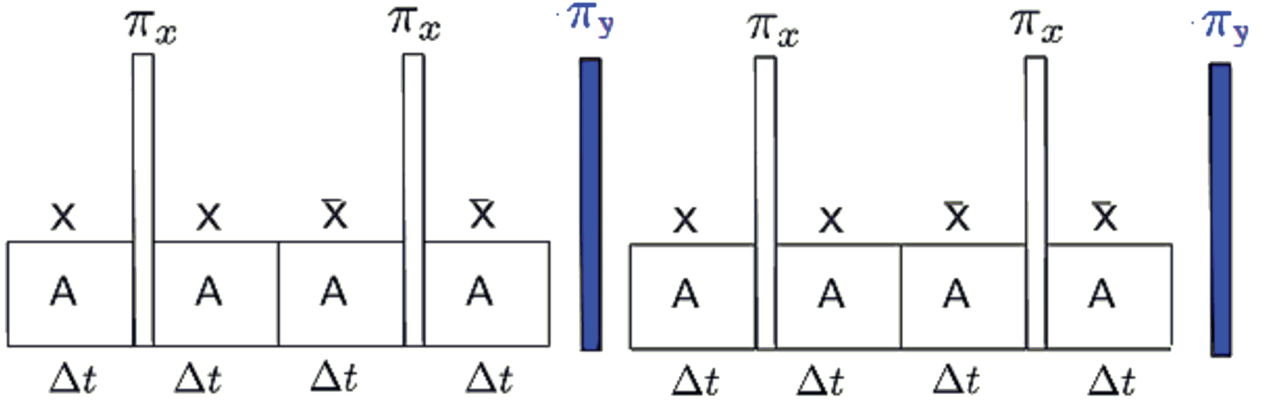}
\end{center}
\caption{Basic pulse sequence for decoupling. }
\label{fig:sequence}
\end{figure}

The pulse sequence is shown in Fig.~\ref{fig:sequence}.

The newly created Zeeman-like terms along the $y$-direction are orthogonal to the Ising coupling terms. Since we have $\theta|\omega_1-\omega_2|\gg J$, we can use rotating wave approximation to reduce the effective Hamiltonian to
\beq
\begin{split}
H_{3eff}&=\frac{\theta}{2}(\omega_1 I_{1y}+\omega_2 I_{2y})+\frac{4}{3}\theta^2 J I_{1y} I_{2y}+O(\theta^3)
\end{split}
\eeq
 that is, if we repeat the procedure $k$ times such that $8k\Delta t\gg \frac{1}{\omega_1-\omega_2}$, then $[e^{-iH_{2eff}8\Delta t}]^k=e^{-iH_{3eff}8k\Delta t}$.
Compare with the original Hamiltonian $H_0$, we have effectively reduced the coupling strength by a factor of $\frac{4}{3}\theta^2$. 
If we can create $|\omega_1-\omega_2|\approx 10^3J$, then $\theta$ can be taken around $\approx \frac{1}{20}$, in this case $\frac{4}{3}\theta^2\approx \frac{1}{300}$, i.e., the coupling strength is reduced by $\approx 300$ times. Further reducing of the coupling strength can be achieved by more iterations of the above procedure. Note that the local terms, such as $\frac{\theta}{2}(\omega_1 I_{1y}+\omega_2 I_{2y})$ can be canceled by Hahn echo pulses, i.e., inserting $\pi$-pulse along the $x$ direction.

This decoupling strategy can be generalized to $N$ qubits with various coupling topologies, which can be easily seen by substituting the Hamiltonian for $N$-qubit system in Eq.(\ref{eq:basicstep}),
\beq
\begin{split}
H_1&=\sum_{j=1}^{N}\omega_j I_{jz}+\sum_{(jk)\in G}J_{jk}I_{jz}I_{kz}+\sum_{j}^N AI_{jx},\\
H_2&=-\sum_{j=1}^{N}\omega_j I_{jz}+\sum_{(jk)\in G}J_{jk}I_{jz}I_{kz}+\sum_{j}^N AI_{jx},\\
H_3&=-\sum_{j=1}^{N}\omega_j I_{jz}+\sum_{(jk)\in G}J_{jk}I_{jz}I_{kz}-\sum_{j}^N AI_{jx},\\
H_4&=\sum_{j=1}^{N}\omega_j I_{jz}+\sum_{(jk)\in G}J_{jk}I_{jz}I_{kz}-\sum_{j}^N AI_{jx},
\end{split}
\eeq

Again by inserting $\pi$-pulses along $\sigma_y$ direction and using rotating wave approximation, this creates an effective Hamiltonian
\beq
H_{eff}^N=\frac{\theta}{2}\sum_{j=1}^{N}\omega_j I_{jy}+\frac{4}{3}\theta^2\sum_{(jk)\in G}J_{jk}I_{jy}I_{ky}++O(\theta^3).
\eeq

The subsequent steps are similar as the ones outlined in the case of a two-qubit system.

Since at every step of the analysis, the precise knowledge of the coupling strength is not required as long as they are small compare to the Zeeman splitting, this decoupling scheme is actually robust to the random fluctuation of the coupling strength $J_{jk}$. For example on the two-qubit system, the coupling strength during the four periods of the basic pulse sequence in Eq.(\ref{eq:basicstep}) can be different from each other---we assume in each period the coupling strength is a constant drawn from a steady distribution, this assumption holds when the frequency of fluctuation is small compare to $\frac{1}{\Delta t}$---in this case, the basic pulse sequence becomes

\beq
\label{eq:basicsteptime}
\begin{split}
H_1&=\omega_1 I_{1z}+\omega_2 I_{2z}+J_1I_{1z} I_{2z}+A(I_{1x}+I_{2x}),\\
H_2&=-\omega_1 I_{1z}-\omega_2 I_{2z}+J_2I_{1z}I_{2z}+A(I_{1x}+I_{2x}),\\
H_3&=-\omega_1 I_{1z}-\omega_2 I_{2z}+J_3I_{1z}I_{2z}-A(I_{1x}+I_{2x}),\\
H_4&=\omega_1 I_{1z}+\omega_2 I_{2z}+J_4I_{1z}I_{2z}-A(I_{1x}+I_{2x}),
\end{split}
\eeq

Keep the terms up to the second order of the average Hamiltonian theory, we obtain the following average Hamiltonian over an interval of $4\Delta t$,
\beq
\aligned
H'_{1eff}&=\frac{J_1+J_2+J_3+J_4}{4}I_{1z} I_{2z}+\frac{A\Delta t}{2}(\omega_1 I_{1y}+\omega_2 I_{2y})\\
&+ \frac{A\Delta t (J_1+3J_2+3J_3+J_4)}{8}(I_{1y} I_{2z}+I_{1z} I_{2y})+\frac{(A\Delta t)^2}{2}(\omega_1 I_{1z}+\omega_2 I_{2z})\\
& +(\frac{J_1+J_2+J_3+J_4}{3}+\frac{J_1-J_2-J_3+J_4}{4})(A\Delta t)^2 (I_{1y} I_{2y}-I_{1z} I_{2z})+O((A\Delta t)^3)
\endaligned
\eeq
which reduces to Eq.(\ref{eq:1eff}) when the four coupling strengths are the same. Again by inserting $\pi$ pulses along $y$-direction, we can flip the signs of the third and fourth terms in $H'_{1eff}$, and creates an effective Hamiltonian
\beq
e^{-iH'_{2eff}8\Delta t}=
e^{-iH'_{1eff}4\Delta t}e^{i\pi(I_{1y}+I_{2y})}e^{-iH'_{1eff}4\Delta t} e^{-i\pi(I_{1y}+I_{2y})}.
\eeq

Here the two $H'_{1eff}$ may have different coupling strength, we assume the coupling strengths for the 4 periods of basic pulse sequences for each $H'_{1eff}$ are $J_1, J_2, J_3, J_4$ and $J_5, J_6,J_7, J_8$ respectively. Then
\beq
\aligned
H'_{2eff}&=\frac{J_1+J_2+J_3+J_4+J_5+J_6+J_7+J_8}{8}I_{1z} I_{2z}+\frac{\theta}{2}(\omega_1 I_{1y}+\omega_2 I_{2y})\\
&+\frac{J_1+3J_2+3J_3+J_4-J_5-3J_6-3J_7-J_8}{8}\theta(I_{1y} I_{2z}+I_{1z} I_{2y})\\
& +\frac{7J_1+J_2+J_3+7J_4+7J_5+J_6+J_7+7J_8}{24}\theta^2 (I_{1y} I_{2y}-I_{1z} I_{2z})\\
&+O((A\Delta t)^3)
\endaligned
\eeq
where $\theta=A\Delta t$. Similar to the case with constant coupling strength, we choose $\Delta t$ such that $\theta|\omega_1-\omega_2|\gg J_i$. By rotating wave approximation the $yz,zy$ and $zz$ coupling are effectively averaged out as they do not commute with $\omega_1 I_{1y}+\omega_2 I_{2y}$, only $yy$ coupling remains, whose strength is reduced by the order of $\theta^2$ comparing to the original coupling strength. The generation to N-qubit system is straightforward, similar to the case with constant coupling strength.

This decoupling pulse sequence can also reduce the dephasing effect caused by the environment. 
Suppose each qubit in the system is coupled to environment where the coupling Hamiltonian is modeled as
\mbox{$H_{SB}=\sum_k \hbar \sigma_z(g_k b^{\dagger}_{k}+g^{\dagger}_{k}b_k),$} where $b_k^\dagger, b_k$ are bosonic operators for the $k$th field mode of the environment, characterized by a generally complex coupling parameter $g_k$\cite{Viola1998}. The $\pi$ pulses along the $x$ and $y$ directions in our decoupling scheme also average out the net effect of $H_{SB}$.

\begin{figure}[bt]
\begin{center}
\includegraphics[width=\columnwidth]{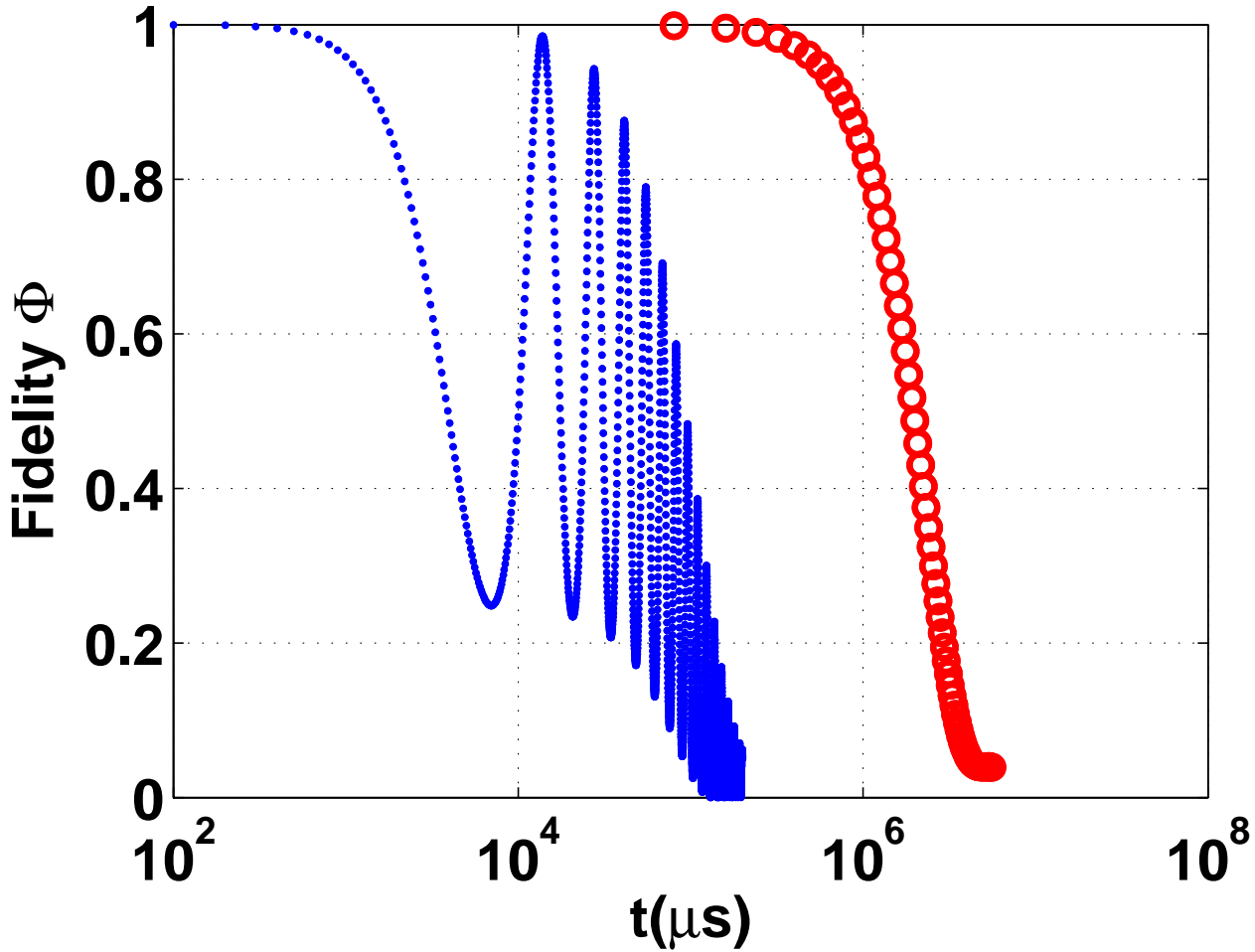}
\end{center}
\caption{(Color online) The simulation was on a square lattice with four qubits. The coupling strengths between these qubits are fluctuating at each instant of time: the coupling strengths between neighboring qubits have mean equal to $J=[17.3, 17.9, 18.5, 19.2]$Hz, and variance equal to the $\bar{J}=9$~Hz; the coupling strength between diagonal qubits have mean equal to $J=[6.1,6.6]$Hz, and variance equal to $3$Hz. The Zeeman splitting of the four qubits are $[62.8, 95.9, 120.1, 153.18]$kHz respectively. $\Delta t=10^{-7}$, $\theta=\frac{1}{20}$, $A\approx 2\pi 8\times 10^3=8 kHz$. The initial state of the qubit is $\frac{|0\rangle+|1\rangle}{\sqrt{2}}$. The blue line shows the fidelity without the global pulses, the red circles show the fidelity with the application of the decoupling pulses.}
\label{fig:numerical}
\end{figure}

Finally, we present a numerical simulation illustrate the effects of our decoupling strategy. As shown in Fig.~\ref{fig:numerical}, the vertical axis represents the fidelity of $U$ with respect to the identity operator, where fidelity measurement between two unitary operators $U_1$ and $U_2$ is defined as
\beq
\phi=\frac{|tr(U_1U_2^{\dagger})|^2}{|tr(U_2U_2^{\dagger})||tr(U_1U_1^{\dagger})|)};
\eeq

Other fidelity measures, for example, the average gate fidelity\cite{Nielsen2002,Fortunato2002}, can also be used, which is equivalent in our case. The simulation was done with 4 spins on a square lattice with one iteration of the pulses. Assume the lattice is put at a vicinity of a dysprosium micro-magnet with length of of $400\mu m$, width $4\mu m$ and height $10\mu m$, which can generate a field gradient of $\partial  B/\partial z=1.4 T\mu m^{-1}$\cite{Ladd2002}. In addition a large homogeneous field $B_0$ of $\sim7$T is superposed. The distance of two neighbor nuclei spin is about $\sim 1nm$, for the simulation we take the distance as $1 nm$ and the nuclear spin as $^{29}$Si\cite{Ladd2002}. If the magnetic field gradient is put along the direction of $y=\frac{\sqrt{3}}{3}x$, then the Zeeman splittings for the $4$ spins are $62.8$kHz, $95.9$kHz, $120.1$kHz, $153.18$kHz respectively. The secular component of the dipolar Hamiltonian which couples the $i$th spin to the $j$th spin is written\cite{Abragam}
$$H_{ij}=\frac{\mu_0}{4\pi}\gamma^2 \hbar^2\frac{1-3\cos^2\theta_{ij}}{r^3_{ij}}I^i_zI^j_z=J_{ij}I^i_zI^j_z,$$
 where $r_{ij}$ is the length of the vector connecting the spins and $\theta_{ij}$ is its angle with the applied field. With chosen parameters, the coupling strengths between adjacent spins are $\sim 17.3$HZ, and $\sim 6.1$Hz between diagonal spins. Due to vibration of atoms, the couplings strength can be fluctuating, so we assume the coupling strengths at each instant of time follow independent normal distributions with mean values $J=[17.3, 17.9, 18.5, 19.2, 6.1, 6.6]$Hz and variance equal to $10$~Hz for the couplings between adjacent spins and $5$Hz for the coupling between diagonal spins. $\Delta t$ is taken to be $10^{-7}$s and $A\approx 8$kHz, $\theta=\frac{1}{20}$. It can be seen that the global pulses reduce the decoherence rate by about two orders of magnitude in one iteration. Note that the results does not depend much on the precise numerical values, as long as the condition $\omega_i\gg J_{jk}$, $|\omega_j-\omega_k|\gg J_{jk},\forall i,j,k$ is satisfied.

\section{Conclusion}
We presented a method that reduces the Ising coupling strength of register qubits using global pulses. This can be used to reduce the residual coupling of quantum memories, where selective addressing may be hard or undesirable. The advantage of these global pulses is that the number of pulses does not grow with the number of qubits, i.e., the complexity of these pulses is only O(1). This opens new directions for using global pulses for decoupling.
\section{Acknowledgements}

H.Yuan acknowledges the financial support from RGC of Hong Kong(Grant 538213).

\end{document}